# On the characterisation of a hitherto unreported icosahedral quasicrystal phase in additively manufactured aluminium alloy AA7075

S.K. Kairy[a], O. Gharbi[a], J. Nicklaus[a], D. Jiang[a], C.R. Hutchinson[a], and N. Birbilis[a,*]

[a]Department of Materials Science and Engineering, Monash University, Clayton, VIC-3800, Australia.

*nick.birbilis@monash.edu

**Abstract**

Aluminium alloy AA7075 (Al-Zn-Mg-Cu) specimens were prepared using selective laser melting, also known as powder bed fusion additive manufacturing. In the as-manufactured state, which represents a locally rapidly solidified condition, the prevalence of a previously unreported icosahedral quasicrystal with 5-fold symmetry was observed. The icosahedral quasicrystal, which has been termed $\nu$-phase (nu-phase), was comprised of Zn, Cu and Mg.

The 7xxx series aluminium (Al) alloys based on the Al-Zn-Mg(-Cu) system are undoubtedly one of the most significant alloy systems of the last century [1-4]. Such Al-Zn-Mg(-Cu) alloys can be considerably age-hardened by precipitates based on the $\eta$-MgZn$_2$ phase (and its close variants), and thus 7xxx series Al-alloys have therefore been the archetypical high strength alloy that has served as the structural material for passenger aircraft [1-8]. The evolution of 7xxx series Al-alloys has been evolving over the past seven decades, with contemporary variants having a higher Zn concentration than early variants (for higher strength) and a higher Cu content (for the control of intergranular corrosion) [9-14]. The industrial production of 7xxx series Al-alloys requires careful control of casting conditions to avoid hot tearing, followed by very detailed (and manufacturer proprietary) thermomechanical processing steps to produce the final desired properties [1-2, 10]. The properties of age-hardenable Al-alloys are extremely sensitive to thermal exposure, and therefore the joining of age-hardenable Al-alloys is complex; requiring bolts in the case of thick sections, and, the numerous - if not infamous - rivets that are the necessary (but not welcome) hallmark of aircraft construction.

Owing to the complexity of joining and producing complex shapes with age-hardenable aluminium alloys, the commercial advent of additive manufacturing permits the design and production of complex components in net-shape. The ability to produce high strength and light weight components from commodity materials (such as Al-Zn-Mg-Cu) provides the impetus for exploring the additive manufacturing of alloys such as Al-alloy AA7075 [15]. The additive manufacturing of AA7075 has been previously by Martin et al., who emphasised the unique aspects of solidification as they pertain to the selective laser melting (SLM) process, and the application of such a process to alloys of compositional complexity. It is clear from the broader recent work employing additive manufacturing to produce metallic alloys, that there is unexplored, if not wonderful, prospects for developing unique alloys and microstructures. Herein, the purpose of the broader study undertaken was to produce AA7075 in net shape with comparable properties to its wrought and thermomechanically processed counterpart. In this pursuit, a number of production variables were explored, with the express aim of producing a dense (pore free) alloy with strengthening achieved upon cooling (as opposed to the conventional aging derived from heating a previously solutionised alloy). In the course of this research, a previously unreported nanoscale icosahedral quasicrystal with 5-fold symmetry was observed in the 'as produced' condition, and was termed $\nu$-phase (nu-phase).

In the study herein, AA7075 powder was supplied by LPW Technology, with a powder size distribution of ~20-63 µm. Selective laser melting (SLM) was carried out using a Concept Laser M-Lab Cusing-R, and specimen production employed a laser power of 95 W, laser scan speed of 200mm/s, layer thickness of 25µm and a hatch distance of 80 µm. The SLM process was carried out in argon purged atmosphere, producing cubes of 15 x 15 x 15 mm upon an aluminium build plate. The elemental compositions of the AA7075 powder and the SLM AA7075 were determined using Inductively Coupled Plasma Atomic Emission Spectroscopy (ICP-AES) by an accredited laboratory (Spectrometer Services, VIC. Australia). The composition of the AA7075 powder was (in wt. %) Al, 4.7 Zn, 2.13 Mg, 1.3 Cu, 0.09 Si, 0.13 Fe, 0.2 Cr. The composition of the final SLM AA7075 cubes was (in wt. %) Al, 3.62 Zn, 1.85 Mg, 1.31 Cu, 0.1 Si, 0.13 Fe, 0.2 Cr. It is evident that there was some loss of both Zn and Mg in the SLM process.

The preparation of specimens for transmission electron microscopy (TEM) was carried out by mechanically thinning alloy samples to 200 µm thickness, followed by punching them to 3 mm discs and finally ion milling to get electron transparent regions using Gatan 691 precision ion polishing system at -100˚C. An FEI® Tecnai G2 T20 and an FEI® Tecnai G2 F20 S-TWIN FEG TEM, both operated at 200 kV, were used for the imaging, electron diffraction and energy-dispersive X-ray spectroscopy (EDXS). An FEI® Titan[3] 80-300 fitted with two (TEM and STEM) CEOS spherical aberration correctors operating at 300 kV was used for aberration corrected bright field imaging. Prior to TEM imaging, specimens were plasma cleaned in an Ar and $O_2$ gas mixture for 2 min using a Gatan® Solarus 950 advanced plasma system. A Bruker X-flash X-ray detector was employed for energy dispersive x-ray spectroscopy (EDXS), and analysis was performed using Bruker Esprit 1.9 X-ray software.

The nanostructure of SLM AA7075 is shown in **Figure 1**, indicating a high density of nanometre length scale particles in the $\alpha_{Al}$-matrix. Such particles formed from solidification during SLM production are intentionally referred to as 'particles' herein, to disambiguate from 'precipitates' that form during heating. The precipitates in conventionally produced and wrought AA7075 are $\eta$-$MgZn_2$ phase (with the possibility of the $\eta$' precursor phase, or the manifestation of $Mg(Zn,Al,Cu)_2$ phase arising from extended heat treatments or Cu rich alloy variants [11, 16]), such precipitates are typically similar in size to the particles observed in **Figure 1**.

A closer inspection of a characteristic 'particle' is given by high magnification BF-TEM in **Figure 2**, in conjunction with corresponding nano-beam diffraction patterns collected from the same particle in three different zones axes. It is evident that the particle in **Figure 2** exhibits characteristic 2-fold, 3-fold and 5-fold symmetries of an icosahedral quasicrystal. In fact, such analysis repeated on numerous particles (and such particles representing the majority of the high particle density in SLM AA7075) indicate the prevalence of an icosahedral quasicrystal phase that forms upon the rapid solidification associated with the aforementioned SLM process.

In order to determine the local chemistry associated with $\nu$-phase, Scanning TEM coupled with EDXS was carried out on a particle colony, as shown in **Figure 3**.

It can be observed (**Figure 3**) that particles are associated with complex compositions that have not previously been reported in the AA7075 alloy system (and which were unable to be indexed using x-ray diffraction). It is noted that the composition of $\nu$-phase includes, in the highest proportions, Zn, Cu and Mg. This precise location of the $\nu$-phase is best assessed by the corresponding Zn-map, which then highlights the co-occupancy of Cu and Mg. The other particles observed in **Figure 3** that are namely associated with Mg-Si and Al-Cu-Fe are synonymous with previously identified compounds (such as $Mg_2Si$ and $Al_7Cu_2Fe$). It is conceded that the resolution and qualitative nature of EDXS do not permit the speculation of the stoichiometry of $\nu$-phase, which is future work.

Employing high resolution Bright Field Scanning TEM (BF-STEM) and aberration correction, direct imaging of the icosahedral quasicrystal structure was possible (**Figure 4**). The ability to directly image complex crystal structures directly using contemporary field-emission Scanning TEM permits the unambiguous structural analysis of nanoscopic phases in metallic alloys – the combination of direct imaging, the Fourier transform of STEM images, and the utility of conventional selected area diffraction, was able to reveal the lattice parameters of $\nu$-phase (seen in **Supplementary Figure A**). An extensive survey of the literature, commencing with the elegant and seminal work of Mondolfo [17] dedicated to Al-Zn-Mg alloys, through to recent works, did not reveal the presence (empirically or computationally) of $\nu$-phase, structurally or compositionally, in prior studies. Whilst not elaborated herein, CALPHAD approaches including ThermoCalc® and PANDAT® did not predict $\nu$-phase under any simulated conditions.

A salient point that merits comment, is that whilst 7xxx series Al-alloys have been in service in critical of applications for decades, there is no prior evidence of the $\nu$-phase reported herein. This reveals that the complexity of physical metallurgy - even for well-studied systems - continues to present itself and is abetted by advanced manufacturing and characterisation techniques. In a similar vein, a new low temperature phase in the age hardening sequence of 7xxx series Al-alloys was only recently revealed using advanced synchrotron characterisation [18], suggesting just how little is known about the engineering of inorganic materials as the next industrial revolution emerges.

**Acknowledgements**

The images herein were collected at the Monash Centre for Electron Microscopy (MCEM), aided by fruitful discussions with M. Weyland. We gratefully acknowledge support by Woodside Energy.

**Figures:**

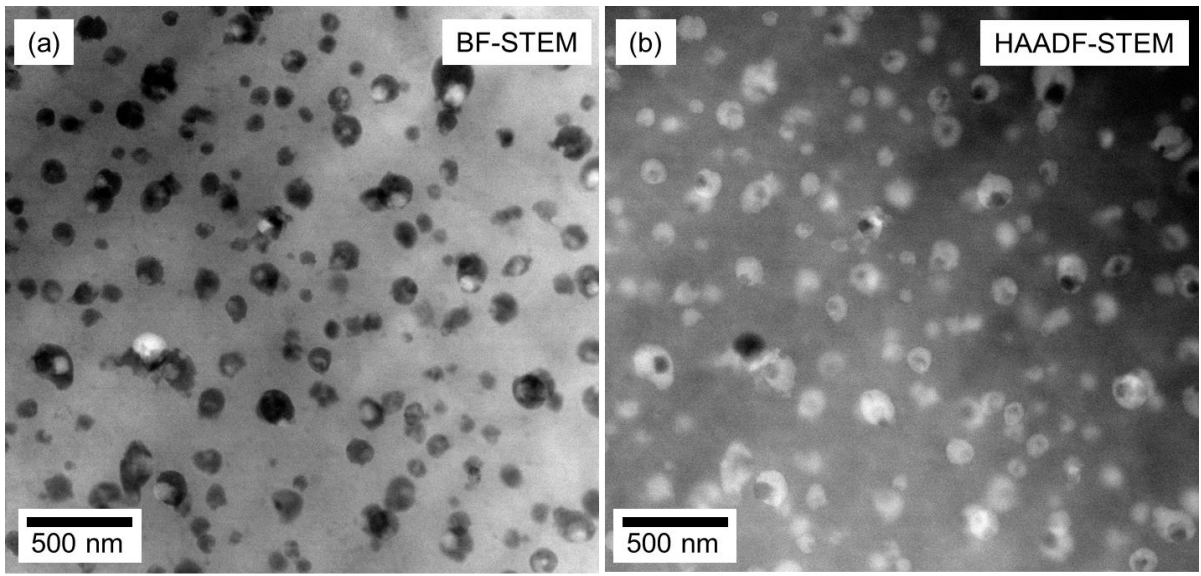

**Figure 1**. (a) BF-STEM and (b) HAADF-STEM images of nanometer scale particles in the $\alpha_{Al}$-matrix of SLM AA7075.

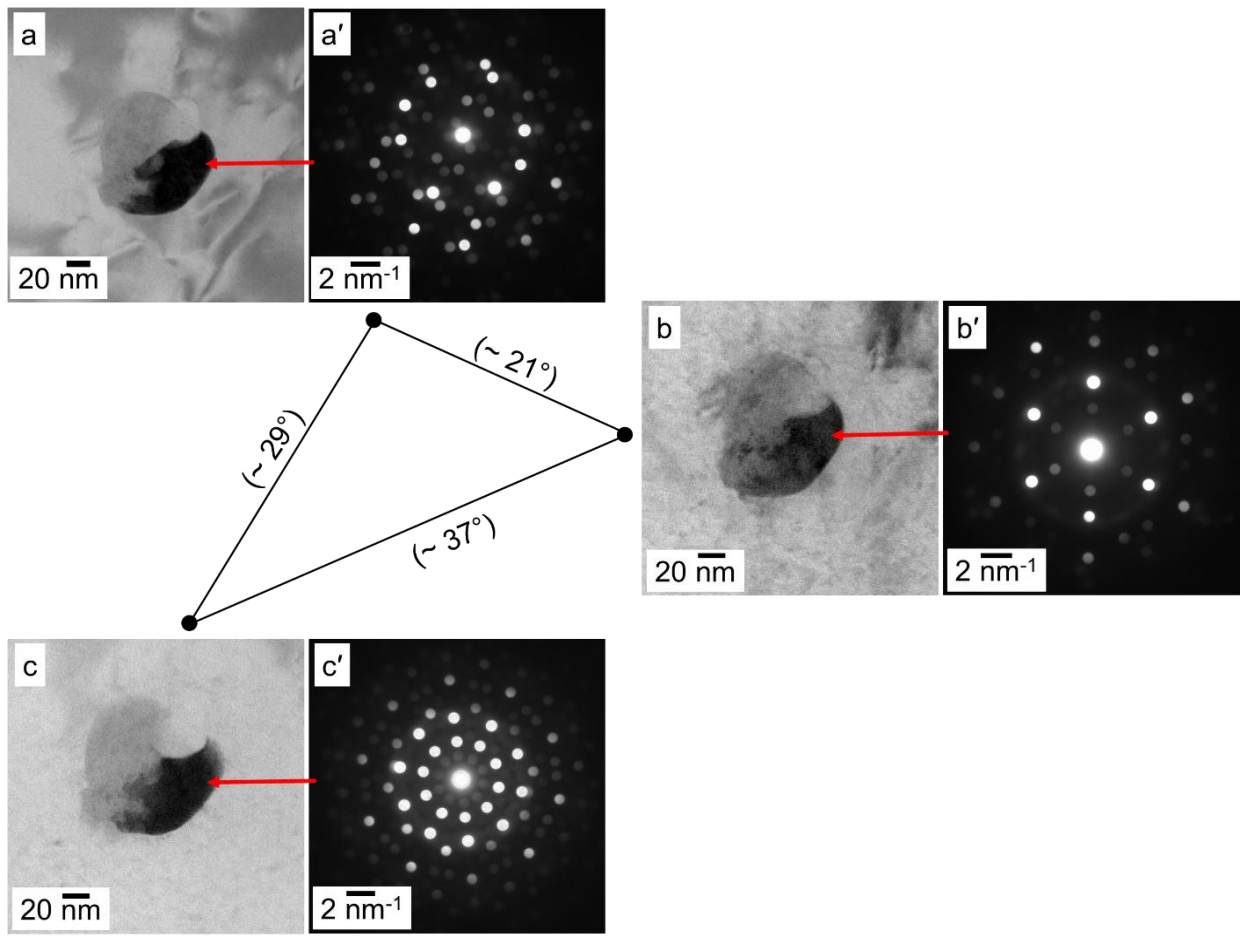

**Figure 2.** (a-c) High magnification BF-TEM images of an identical location containing nanometer scale particles in the $\alpha_{Al}$-matrix of SLM AA7075. The images were collected upon tilting the same particle, exhibiting strong diffraction contrast (red arrow), in all three zone axes. (a′-c′) Nano-beam diffraction patterns collected from the same particle in three different zones axes, exhibiting characteristic 2-fold, 3-fold and 5-fold symmetries of an icosahedral quasicrystal, respectively.

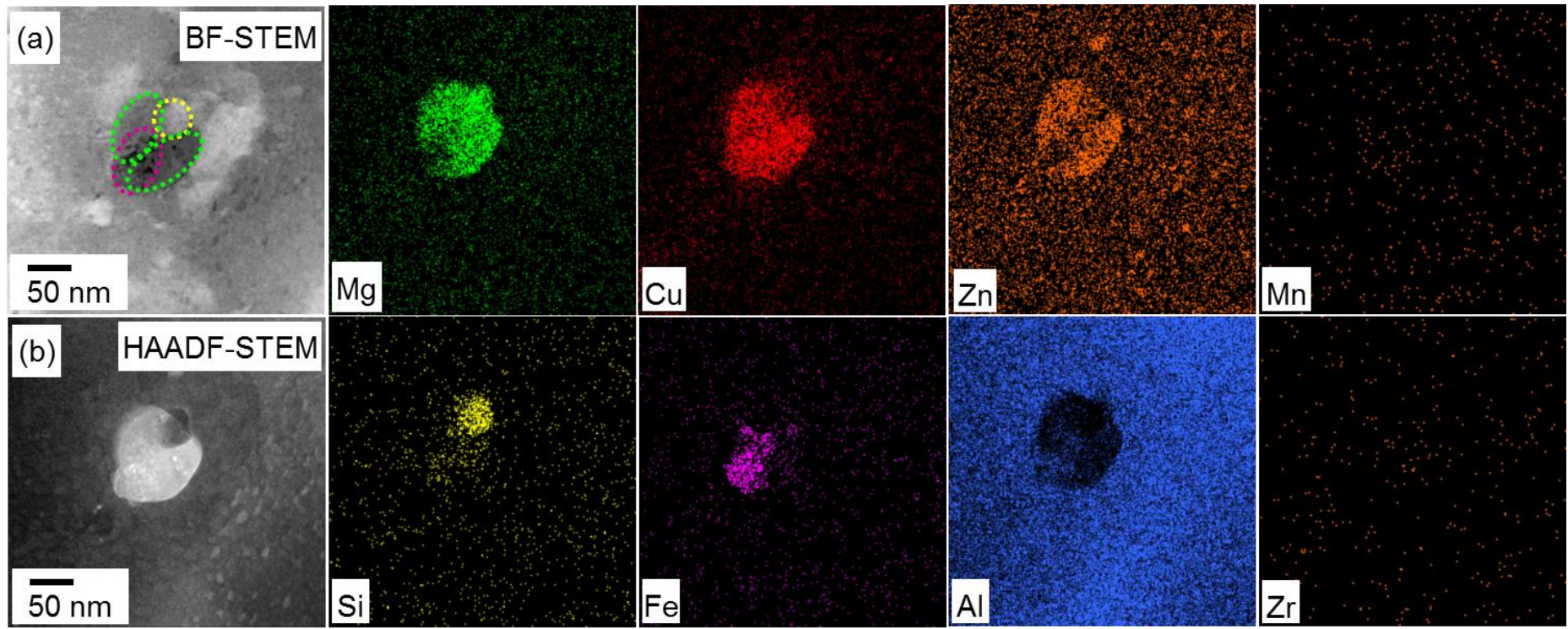

**Figure 3**. (a) BF-STEM and (b) HAADF-STEM images, of the identical location in Figure 2, containing nanometer scale particles in the α$_{Al}$-matrix of SLM AA7075. The EDXS maps corresponding to the images are presented. Individual particles enclosed in green, yellow and purple dotted lines in (a) are enriched in Cu-Mg-Zn, Mg-Si and Al- Cu-Fe, respectively.

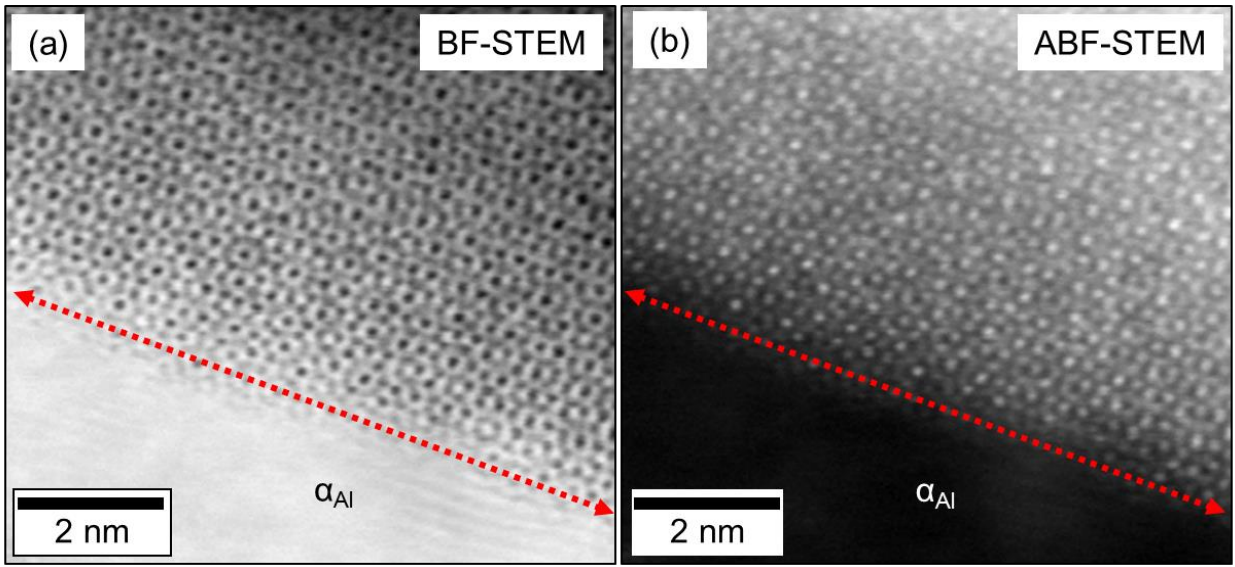

**Figure 4**. (a) BF-STEM and (b) ABF-STEM images of an icosahedral quasicrystal particle, along a 5-fold symmetrical zone axis, in the $\alpha_{Al}$-matrix of SLM AA7075. Dotted red arrows indicate the quasicrystal particle-matrix interface.

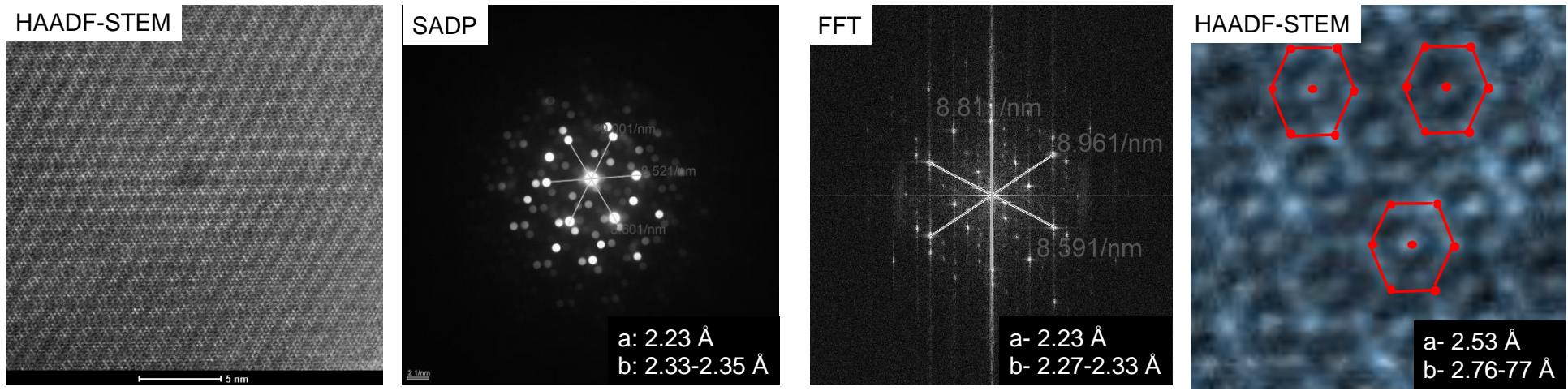

**Supplementary Figure A**. Determination of lattice parameters based on analysis of the 2-fold symmetry of the quasicrystals shown in **Figure 1**.